\documentclass[12pt]{article}

\usepackage[latin1]{inputenc}

\usepackage[english]{babel}

\usepackage{epsfig}
\usepackage{amsmath}
\usepackage{amssymb}
\usepackage{amsfonts}
\usepackage{bbold}

\def\s{\sigma}

\def\o{\over}

\def\a{\alpha}

\def\c{\gamma}
\def\b{\beta}
\def\d{\delta}

\def\u{\omega}

\def\.{\cdot}
\def\p{\partial}

\def\+{\bigoplus}

\def\({\left(}
\def\){\right)}
\def\[{\left[}
\def\]{\right]}
\def\l.{\left.}
\def\r.{\right.}

\def\o{\over}
\def\<{\left\langle}
\def\r|{\right|}
\def\>{\right\rangle}
\def\l|{\left|}
\def\nn{\nonumber\\&&}

\def\fns{\footnotesize}

\def\beq{\begin{equation}}
\def\eeq{\end{equation}}
\def\bea{\begin{eqnarray}}
\def\eea{\end{eqnarray}}
\def\nn{\nonumber \\ &&}

\def\ber{\begin{array}}
\def\eer{\end{array}}

\newcommand{\newsection}[1]{\section{#1}\setcounter{equation}{0}}

\begin{document}
\begin{titlepage}
\noindent

July, 2016\hfill 
\vskip 1in
\begin{center}

{\large \bf Renormalizable theories with symmetry breaking}
\footnote{The  present manuscript contains a partial  revision and update of  the results of a work done in collaboration with A. Rouet and R. Stora in 1974-5 and  published after six years in \cite{brs}. The author of the present manuscript has tried to keep unchanged the largest possible part of the original paper and takes full responsibility for any  mistakes.}
\vskip 0.3in

Carlo M. Becchi \footnote{E-Mail: becchi@ge.infn.it}
\vskip .2in

{\em Dipartimento di Fisica, Universit\`a di Genova,\\
Istituto Nazionale di Fisica Nucleare, Sezione di Genova,\\
via Dodecaneso 33, 16146 Genova (Italy)}
\end{center}
\begin {abstract}
 The description  of  symmetry breaking proposed  by  K.  Symanzik  within  the
framework  of  renormalizable  theories  is  generalized  from  the  geometrical  point of  view.
For  an  arbitrary  compact  Lie  group, a soft breaking  of arbitrary  covariance, and an arbitrary
  field  multiplet,  the expected  integrated Ward  identities  are  shown  to hold to all orders  of
renormalized  perturbation theory provided  the  Lagrangian  is  suitably  chosen.  The  corresponding 
local Ward identity which  provides the Lagrangian version  of current  algebra
through the coupling  to  an external,  classical,  Yang-Mills  field,  is  then proved  to hold  up to
the  classical  Adler-Bardeen anomaly  whose general  form is written  down. The BPHZ
renormalization  scheme  is  used  throughout  in  such  a  way  that the  algebraic structure
analyzed  in  the present  context may  serve  as  an  introduction to  the study of  fully  quantized
gauge  theories.
\end{abstract}

\vfill

\end{titlepage}
\setcounter{footnote}{0}

\section{Introduction}

Besides  the well-known  relevance  of  broken  symmetries  to  elementary
particle  physics,  further investigations  of  renormalizable models  exhibiting
broken  symmetries are amply  justified  by  the  present understanding  of
gauge  and  even super-gauge theories.  The  present  status of the  subject  is
well represented  by  K.  Symanzik's  1970 Carg\'ese  lectures where the  fundamental phenomena  are discussed  \cite{sym}.  Since  much  of  the  structure  analyzed
there can also be  found  in  the study of  gauge theories,  it  is of interest to
complete  Symanzik's  analysis  from  the  algebraic  point of  view,  both in
considering an  arbitrary  compact  Lie group  as  describing the symmetry
to  be  broken,  and  in  analyzing the perturbative  analog of  current  algebra,
namely the coupling  with  an external  Yang-Mills  field, which,  as  is well-known, leads  to  the  definition  of  the celebrated  Schwinger-Bell-Jackiw-Adler-Bardeen  anomaly \cite{adl}.  In  fact,  although this program  had  been  announced 
 by  K.  Symanzik,  
  \cite{sym},  it  has
not  been  carried  out  until  now  following  the  stimulation  \cite{brs1}  provided by
the  advent  of gauge  theories. Also,  it  seems  that  recent  progress  in  renormalization 
 theory  has  allowed a more tractable  general  treatment  than  the
techniques known in  1970  would have permitted. Most of the present analysis
relies on  general properties  of the perturbative series which stem from locality
and  power counting,  as  summarized  by  the  renormalized action principle
of  Lowenstein and  Lam  \cite{lol} whose  detailed  form  is  one  of the highlights
of  the  Bogoliubov-Parasiuk-Hepp-Zimmermann  \cite{zim} renormalization
scheme.  Within  this framework,  explicit  bases  of local operators of  given
dimensions are constructed  \cite{gol},  together  with  the linear relations  connecting 
operators  with  different dimensions (the so-called Zimmermann
identities).  To  avoid  inessential  technical complications, we shall  only
treat  cases  in  which  no  massless  field  is  involved.  Our  analysis  can,
however, be  extended  without  essential  modifications  to  a  wide  class  of
models  involving  massless  fields exploiting  Lowenstein's  and  Zimmermann's extension  of  BPHZ  renormalization  scheme  \cite{lzc}.  Whereas
these  elementary tools,  which  are best  exploited  by  means  of  a  repeated
application of  the  implicit  function  theorem  for  formal  power  series   \footnote{\label{1}For a brief summary see \cite{brs1} (b), Appendix II.},
suffice  to  solve  most  of the algebraic  problems at  hand, the  elimination of
some  possible  anomalies  is  occasionally performed  by  looking  more
deeply  into  the behavior  of  the theory under  scaling transformations,
which  provides  some  new  non-renormalization type  statements similar
to  that  which  leads  to  the non-renormalization  of  the Adler-Bardeen
anomaly coefficient  \cite{ls}\cite{b}.
This article  is  divided  into  two  main  parts:  Section 2  is devoted  to  the
proof  of the  integrated Ward  identity  which  expresses  symmetry breaking
for  an  arbitrary  compact  Lie  group,  with  an  arbitrary  dimension  $(  <  4)$
and covariance. Section  3  is devoted  to  a  discussion  of  'current  algebra'
which,  in  the  present  framework  amounts  to  the proof  of  a local  Ward
identity,  in  the  presence  of  an external  Yang-Mills  field, and  leads  to  the
definition  of  the  Adler-Bardeen  anomaly.
A number  of appendices  are  devoted  to  the  treatment of  some  technical
questions,  among  which  the  elimination  from  the  integrated  Ward
identities  of  algebraically  allowed  anomalies consistent  with  power
counting,  and  details about the cohomology   of  the gauge  Lie  algebra
associated  with  the symmetry group  (i.e.  the  Wess-Zumino \cite{wz} consistency
  conditions).
\section{  Broken global  symmetries }
This chapter is devoted to the proof of the perturbative renormalizability of a generic model built on a set of quantized field variables and characterized by a softly broken invariance under field transformations belonging to a compact Lie group. We shall systematically use a functional formulation in which, e.g., Green's functions are obtained as functional derivatives of their functional generator,  and the classical Lagrangian is a local field functional.\footnote{Thus quantized fields can also be  interpreted as  functional variables.} The need of describing local operators such as e.g. the terms breaking the invariance  of the classical Lagrangian,  requires the introduction, together with the quantized fields, of further functional variables, that we call  external fields,  coupled to the relevant operators. 

It might be useful to shortly remind the general properties of a  perturbatively renormalizable theory, in particular, in the chosen, regularization independent, framework based on the BPHZ scheme.

First of all, the perturbative construction is based on the Feynman diagram expansion.\footnote{
In Feynman amplitudes quantized  fields propagate while external ones do not. In the functional
 formalism to every quantized field one associates a further functional variable called the field source which
 plays the role  of Legendre conjugate variable  to the quantum field. The Legendre transform of the 
classical action is the functional generator of the tree-approximation Feynman diagrams.} The kernel of Feynman's construction is the calculation of  1-particle irreducible (1-P.I.) diagrams amputated of their external legs. 
 Their functional generator is called the {\it effective action} and denoted by $\Gamma(\varphi)$. A $n$ loop 1-P.I.  diagram corresponds to an amplitude proportional to $\hbar^n$, thus $\Gamma(\varphi)$ is a  formal power series in $\hbar$. In the classical limit Feynman diagrams correspond to tree diagrams, those without loops, and $\Gamma(\varphi)$ corresponds to the classical action. In our scheme in the fully quantized limit, not only Green's functions, but also many important quantities, as e.g. Lagrangian parameters,  are formal power series in  $\hbar$.

Renormalizability is based on power counting. A canonical (power counting) dimension is associated with any field, in particular, in the case of quantized fields, this dimension is determined by the maximum derivative degree of the free, bilinear part of the Lagrangian, or else,  of the higher derivative part of the  wave operator which is assumed non-degenerate. As is well known the short distance behavior of the  causal Green's function, the propagator, is  determined by the dimension of the corresponding fields. The general  necessary condition for renormalizability is that the canonical dimension of the Lagrangian,  also including the contribution of derivatives,  should not exceed four.\footnote{Note that e.g. in the case of a massive vector field where gauge invariance is broken by a mass term the higher derivative part of the  wave operator is degenerate due to gauge invariance and the model is not renormalizable. } 

In many important cases, once the classical Lagrangian is given, one builds Feynman diagrams, and hence the effective action, computing  suitably regularized 1-P.I.  diagrams, so avoiding divergent results. A clever choice of regularization my help in preserving symmetry properties of Green's functions. In reality this works well in some cases, while it is not a universal method. Choosing BPHZ subtraction method we  have a systematic construction of Green's functions, but symmetry might be  broken by loop corrections. The aim of the present paper is to show how symmetry can be restored even in the BPHZ  framework and hence independently of regularization. Zimmermann's subtraction method  associates with every vertex in a Feynman diagram a quantized field dependent monomial  $M(\varphi)$ equipped with the  prescription that the non trivial (sub)-diagrams containing the vertex should be subtracted at zero momenta of the external legs together with their Taylor expansion up to total  dimension $\d\geq {\rm dim} M$.   The operator corresponding to the `subtracted' monomial is denoted by $N_\d[M]$ and $\d$ is called Zimmermann's index.\footnote{The great advantage of Zimmermann's subtraction method is its precise definition and the identification of  complete bases of local operators with  well defined power counting behavior. In spite of the  very careful and  detailed form of the original  formulation, it is possible to show that essentially the same properties are obtained by other methods, e.g. renormalization group evolution equations, in which extra subtractions correspond to stronger initial conditions.\cite{bb} } For the terms of the effective Lagrangian ${\cal L}_{\rm eff}$, which contains all the prescriptions for the Green function construction, an $N_4$ subtraction  is understood. A second basic point is Lowenstein-Lam's quantum action principle  \cite{lol} according to which the variation of  $\Gamma (\varphi) $ under infinitesimal parameter and field transformations corresponds to the {\it insertion}  into $\Gamma(\varphi)$ of a (possibly integrated) local vertex whose Zimmermann's index is the maximum canonical dimension of the variation of ${\cal L}_{\rm eff}$, four in our case. The insertion of the vertex $V(\varphi)$ into $ \Gamma(\varphi)$  corresponds to the introduction
into every 1-P.I. diagram contributing to  the expansion of  $\Gamma(\varphi)$ of a 
further  local vertex which is specified by the form of $V$. In general $V$  may depend on both external and quantum fields. The insertion of the vertex $V(\varphi)$ into $ \Gamma(\varphi)$ is denoted by $V\Gamma(\varphi)$ which is a new $\hbar$ formal power  series   valued functional satisfying the equation $$V\Gamma(\varphi)=V(\varphi)+O(\hbar V)$$ where $V(\varphi)$ is interpreted as a local functional  and $O(\hbar V)$ lumps the contributions of the non trivial loop diagrams together.  There are exceptional situations which correspond to operators, either independent, or linear in the quantized fields. In these cases the corresponding vertices cannot be inserted into 1-P.I. diagrams and hence the insertion of an exceptional operators into $\Gamma(\varphi)$ is purely additive $V\Gamma(\varphi)=V(\varphi)$.

A general quantization condition for any system is the stability of its dynamics under infinitesimal changes of parameters and consistent deformations of symmetry conditions.
Dealing with $\hbar$ formal power series, implicit function theorem says that the mentioned stability properties are guaranteed if they hold true at the zeroth order, that is, in the classical theory. Thus  if, e.g., we want  to  construct a  perturbation theory  for  which  a  particle  interpretation  exists,  we must  assume that there exists an invertible change of variables between  the parameters of the classical  Lagrangian  and  the physical ones.
In general  we shall precede the analysis of any quantum property by a discussion of the classical case and of its stability under   change of  parameters and symmetry conditions.  We shall denote by an upper ring the classical quantities with the exception of the  the classical Lagrangian density functional/operator   ${\cal L}$. Thus we have the functional equation
 $$\mathring\Gamma=\int dx\ {\cal L},$$ and the operator equation$$\mathring {\cal L}_{\rm eff}=N_4[{\cal L}].$$ 


\subsection{  The  Classical  Theory}
The general  situation is  as  follows:  G is  a compact  Lie  group,  ${\cal G}$   its Lie
Algebra:  ${\cal G}={\cal S}  +{\cal A}$, ${\cal S}$ semi-simple,  ${\cal A}$  Abelian.  ${\pmb \varphi}$   
 is a field  multiplet belonging  to  a fully  reduced  finite-dimensional  unitary  representation
$D$ of  $G$,  $d_{\varphi}>0$ is the  canonical dimension of ${\bf \varphi}$.  Given $X\in {\cal G}$ the corresponding infinitesimal
transformation  of  ${\pmb \varphi}$     is:
\beq \d_X {\pmb \varphi} =-\mathring {t}(X)   {\pmb \varphi} \eeq
where $ X  \to \mathring t(X) $   is the  representation of  ${\cal G}$     induced  by $D$.

Let  ${\pmb \b}$ be a classical  field
 to  which  is  assigned dimension  $d_\b <4 $,  belonging
to  a  multiplet  characterized  by  another  representation ${\cal D}$     of  $G$  (finite
dimensional,  fully  reduced,  unitary,  with  no  identity  component) and
\beq X\to \mathring \theta (X) , \quad   X\in  {\cal G } ,   \eeq  be 
the  corresponding representation  of  ${\cal G}$.
The symmetry  $G$  will  be said  to  be  broken  with  dimension  $4  -  d_\b$,
covariance  $\mathring{\pmb b}$,  belonging  to  multiplet  ${\cal D}$,  if  there  exists  a  Lagrangian
${\cal L}({\pmb \varphi},{\pmb \b})$  of maximum  dimension  four  invariant  under  the  simultaneous
transformation
\bea&&   {\pmb \varphi} \to {}^g{\pmb \varphi}=D(g^{-1}) {\pmb \varphi,}\\&&
{\pmb \b}+ {\pmb b} \to {}^g({\pmb \b}+ {\pmb b}) ={\cal D}(g^{-1})({\pmb \b}+ \mathring{\pmb b}).
\eea
The classical  field  ${\pmb \b}$  is  introduced  as  an  auxiliary  item,  in  order  to  characterize
  the breaking  described  by the  space-time  independent  $  \mathring {\pmb b}$   according
to  its  dimension,  a  concept  which  is  meaningful  in  the  renormalizable
framework  we have  in  mind. The theory  will  be  truly  renormalizable, i.e.  ${\cal L}({\pmb \varphi},{\pmb \b})$  will  be a  polynomial,  if
\beq d_{\pmb \b} >0,\eeq This  criterion,  introduced  by Symanzik \cite{sym}, leaves  the broken theory  with  an  asymptotic  memory  of  the  initial
symmetry group. However, the  limiting  case
$$  d_{\pmb \b} =0,$$
can also  be  considered since   ${\cal L}({\pmb \varphi},  \mathring{\pmb b})$   is  invariant  under  simultaneous
transformation  of $ {\pmb \varphi}$   and $ \mathring{\pmb b} $  and   is  not,  in  general,  the most general  Lagrangian  which  is  invariant  under  the residual  symmetry  group, namely the
stability  group  $H_{ \mathring{\pmb b} }$  of  $\mathring {\pmb b} $.  Clearly,  the notion  we  have  introduced only
depends  on  the  equivalence  classes  of $D$, ${\cal D}$  and  the  orbit of  $\mathring {\pmb b} $.  We  shall
assume  that in  the  tree  approximations  of  the  corresponding
Green  functions and  for  some  values  of  the  parameters characterizing  ${\cal L}$,  a particle  interpretation  is  possible and  that  there is an  invertible  change  of parameters  between  the  coefficients  of  ${\cal L}$  and those  occurring  in  normalization  conditions  
 through which  masses,  coupling  constants,  etc.,  are
defined.  We shall furthermore  assume  that no  vanishing  mass  parameter
appears  in  the  theory.
When  the  Lagrangian has  a  term linear  in  the  quantized field,  the
particle  interpretation  requires   a  field  translation
$$ {\pmb \varphi} \to {\pmb \varphi} + \mathring {\pmb F}  $$
through which  the  linear  term  is  eliminated.  $\mathring {\pmb F}  $ is  then defined  by:
$$ \p_\varphi  {\cal L} ({\pmb \varphi} ,  \mathring{\pmb b}) \vert_{\varphi= \mathring{F} } =0$$
which  certainly has a  solution  continuous  in  the parameters  of
${\cal L}$  if  the  mass  matrix
$${\cal M} = \p^2_{\varphi\varphi} {\cal L} ({\pmb \varphi} ,  \mathring{\pmb b}) \vert_{\varphi= 
\mathring{  F}}$$
is  non-degenerate.  As  shown  in Appendix   A, $ \mathring{\pmb F} $ is then a covariant function of 
$\mathring {\pmb b} $, and, consequently, the coefficients of the Lagrangian expressed in
terms  of  the translated  fields  are also covariant.  From  now on,  we  shall still denote by $ {\pmb \varphi} $ and $ {\pmb \b} $
the  translated field and  by  $$\tilde {\cal L} ( {\pmb \varphi}, {\pmb \b})\equiv {\cal L} ( {\pmb \varphi}+ \mathring{\pmb F} , {\pmb \b}+\mathring{\pmb b} ). $$

At  this point,  the  action
\beq  \mathring\Gamma({\pmb \varphi}, {\pmb \b}) =\int dx\ \tilde{\cal L}({\pmb \varphi}, {\pmb \b}) \eeq
fulfills  the  {\it integrated}  Ward  identity
\bea W ( X ) \mathring\Gamma({\pmb \varphi}, {\pmb \b})\equiv&&-\int dx\Bigr\{ {\d \mathring{\Gamma}\o\d{\pmb \varphi} } \mathring{t} (X) ({\pmb\varphi}+\mathring {\pmb F})\nn
+{\d \mathring\Gamma\o\d{\pmb \b} } \mathring\theta (X) ({\pmb\b}+ \mathring {\pmb b})\Bigr\}=0,
\quad\quad X\in {\cal G}
\label{sette}\eea
which  expresses  its invariance under  the  infinitesimal transformation
\bea \d_X{\pmb \varphi}&&=- \mathring{t} (X) ({\pmb\varphi}+ \mathring {\pmb F}),\nonumber\\
\d_X{\pmb \b} &&=- \mathring\theta (X) ({\pmb\b}+ \mathring {\pmb b}). \label{pippo}\eea

Now we  note   that  if  we wants  to  construct a  perturbation
theory  for  which  a  particle  interpretation  exists,  it  is  necessary
to  assume the  following:  Let  the Lagrangian  be  written in  the  form
\beq {\cal L } = \mathring {\pmb  C}_\sharp {\pmb{\cal L}}_\sharp =\sum_i \mathring C^i_\sharp  {\cal L}^i_\sharp\label{9}\eeq
where the  $ \mathring C^i_\sharp  $'s  are  numerical coefficients  and    ${\cal L}^i_\sharp$ are all  possible  local
monomials  invariant  under Equation (\ref{pippo}),  consistent  with  the  renormalizability  requirement.  Then 
there  must exist an invertible  change  of  variables between,  on  the one hand, the  $\mathring  C^i_\sharp  $'s  and $\mathring  {\pmb b}$ 
  and, on  the other  hand,  a  set  of physical parameters (masses,  wave  function  normalizations,  
coupling  constants)  occurring  in normalization conditions  imposed  on  $ \mathring{\Gamma}$. These  normalization conditions must be
  consistent with  the symmetry  expressed  by  the  Ward identity, but  not  constrained
by  power  counting.  This implies in  particular that  power  counting  does  not
restrict Equation  (\ref{9})  compared  to  the most  general  solution of  Equation
(\ref{sette}), as  far  as  these  normalization conditions  are concerned
(e.g.  power counting does  not enforce  mass  rules).  Of  course, the  fulfillment  of  normalization  conditions
is  only  necessary  if  a  particle  interpretation  is required, 
the  Ward identity  being sufficient  if  only  a  theory  of  Green's  functions  is aimed  at.

Secondly one  might  object  that  the  prescription  of  the  Ward identity Equation  (\ref{sette})
does  not  seem  to  define the  theory in  a  natural  way  from  the  point  of view
of  power counting:  in a  more  general  scheme  one would    have  a  Ward
identity  with  the  following  structure:
\bea W ( X ) \mathring\Gamma({\pmb \varphi}, {\pmb \b})\equiv&&-\int dx\Bigr\{ {\d \mathring\Gamma\o\d{\pmb \varphi} } (\mathring T (X) {\pmb\varphi}+ \mathring {\pmb F}(X))\nn
+{\d \mathring\Gamma\o\d{\pmb \b} } (\mathring \Theta (X) {\pmb\b}+ \mathring {\pmb b}(X))\Bigr\}=0,\quad\quad X\in{\cal G}
\label{10}\eea
subject  to  the  algebraic constraint
\beq \[W(X),W(Y)\]=W(\[X,Y\]), \quad\quad X,Y\in{\cal G}\label{11}\eeq
thus expressing the  invariance of  $\mathring{\Gamma}$  under the  transformation
\bea \d_X{\pmb \varphi}&&=- \[\mathring T (X) {\pmb\varphi}+ \mathring {\pmb F}(X)\],\nonumber\\
\d_X{\pmb \b} &&=- \[\mathring \Theta (X) {\pmb\b}+\mathring  {\pmb b}(X)\].\eea
with  coefficients constrained by:
\bea\label{13}&&a)\quad \[\mathring T(X),\mathring T(Y)\]  =  \mathring T(\[X,Y\]),  \nn b)\quad \[\mathring \Theta(X),\mathring \Theta (Y)\]  = \mathring \Theta(\[X,Y\]),  \nn c)\quad 
\mathring T(X)\mathring  {\pmb F}(Y) - \mathring T(Y) \mathring {\pmb F}(X) -\mathring {\pmb F}(\[X,Y\])=0, \nn d)\quad \mathring \Theta(X)\mathring  {\pmb b}(Y) - \mathring \Theta(Y) \mathring {\pmb b}(X) -\mathring {\pmb b}(\[X,Y\])=0
,\eea
according  to  which
$$X  \to \mathring T(X) ,$$
$$X  \to \mathring \Theta(X) ,$$
are  representations of  $ {\cal G} $  and  $\mathring {\pmb b}(X) $,  $\mathring {\pmb F}(X) $  are $ {\cal G} $ Lie algebra cocycles  \footnote{For a brief summary of Lie algebra cohomology, in particular the meaning of coboundary and cocycle,  see \cite{brs1} (c), Appendix A }   
with  values in  the  representation  spaces  $E $ and  ${\cal E}$ of  $\mathring T$  and  $\mathring \Theta$ respectively. It  is  shown  in 
Appendix  B that  the  requirements  which allow  a  particle interpretation  and  take
into  account  our  definition  of  symmetry  breaking  put  quite  severe
restrictions  on $ \mathring T (X) $,  $\mathring \Theta(X) $, namely  they  can be  lifted  to  the group  $G$,
and thus,  in particular,  they  are  fully  reducible.  They  can thus  be  obtained
from  representatives  of  their  equivalence  classes  which are related to $ \mathring{t}$,  
$ \mathring\theta$, through  suitable  field  renormalizations:
\bea \mathring T(X)=&&  Z^{-1} \mathring{t}(X)Z,\nonumber \\
\mathring \Theta(X)=&&  Z^{-1} \mathring\theta(X)Z,\eea
It  is  then shown  in  Appendix  B  that $\mathring {\pmb F}(X)$,  is  a  Lie algebra coboundary 
\beq \mathring {\pmb F}(X)= \mathring {t}(X)\mathring {\pmb F}\eeq
for  some  fixed  $\mathring {\pmb F}$,  up  to  invariant  components
$$\mathring  {\pmb F}_\sharp  (X) $$
which  vanish for $X\in {\cal S}  $,  the semi-simple  part  of  ${\cal G}$.  It  is  finally  shown  in
Appendix  B that  $\mathring {\pmb F}_\sharp  (X) \not= 0$  contradicts  the  assumption  that  the  mass
matrix  is  non-degenerate.  Thus  the Ward identity  (Equation (\ref{sette}))  is
actually the most  general  in  the  present context,  since  the Lie algebra  coboundary
structure  of  $\mathring {\pmb b} (X)$:
$$\mathring  {\pmb b}(X)=\mathring{\theta}(X)\mathring  {\pmb b},$$
is  implied by our picture of symmetry breaking.
\subsection {  Radiative Corrections}
The description of radiative corrections  proceeds  via the construction  of
an  effective  dimension  four  Lagrangian
\beq {\cal L}_{\rm eff} = N_4[{\cal L} + \hbar \Delta {\cal L}]\eeq
without  a  term  linear  in the quantized fields,  with  coefficients  formal power series in $\hbar$ such that the 
renormalized Ward identity holds: 
\bea W ( X ) \Gamma \equiv&&-\int dx\Bigr\{ {\d  \Gamma\o\d{\pmb \varphi} } ({t} (X) 
{\pmb\varphi}+ {\pmb F}(X))\nn +
{\d \Gamma\o\d{\pmb \b} } (\theta (X) {\pmb\b}+ {\pmb b}(X))\Bigr\}=0,\quad\quad X\in{\cal G}=0\label{17}
\eea
subject  to  the  algebraic constraints strictly analogous to those given in Equation (\ref{11}).
Equation (\ref{17}) expresses  the  invariance  of  $ {\cal L}_{\rm eff}$
in  the  sense  of  the renormalized
action principle  under the  renormalized transformation
\bea \d_X{\pmb \varphi}&&=-( t (X) {\pmb\varphi}+ {\pmb F}(X)),\nonumber\\
\d_X{\pmb \b} &&=- (\theta (X){\pmb\b}+  {\pmb b}(X)). \eea
which  coincides  with  Equation  (\ref{pippo})  in  the lowest order  in  $\hbar$.
The  analysis performed  in  this  section  will  actually lead  to  the
conclusion  that  there exists  a  quantum  extension  in  which the  almost
naive  Ward identity  holds:
\bea W ( X ) \Gamma({\pmb \varphi}, {\pmb \b})\equiv&&-\int dx\Bigr\{ {\d \Gamma\o\d{\pmb \varphi} } 
\mathring{t}(X) ({\pmb\varphi}+ {\pmb F})\nn +
{\d \Gamma\o\d{\pmb \b} } \mathring{\theta} (X) ({\pmb\b}+ {\pmb b})\Bigr\}=0, \quad\quad X\in{\cal G}\label{19}
\eea
where    ${\pmb F}$  is  determined  by  the requirement that  ${\cal L}_{\rm eff} $  has  no
term linear  in  ${\pmb\varphi}$, and ${\pmb b}$,  which picks up
radiative  corrections due to normalization conditions, has  the  same  stability  group  $H_{\mathring {\pmb b} }$ of
 $\mathring {\pmb b} $. This  is  to  say  that  one can  fulfill  quantum invariance
under  the  renormalized  transformation
\bea \d_X{\pmb \varphi}&&=- \mathring{t} (X) ({\pmb\varphi}+ {\pmb F}),\nonumber\\
\d_X{\pmb \b} &&=- \mathring\theta (X) ({\pmb\b}+ {\pmb b}),\label{20}\eea
as a consequence  of  the  algebraic constraints, Equation  (\ref{11}). It is shown  in
Appendix  B  that the  consistency  conditions (Equations (\ref{13})) on  $t(X)$  and  $\theta(X)$  can be used  to
replace  them  by $\mathring t(X)$  and  $\mathring \theta(X)$ up   to  terms  which  
can be interpreted  as
invariant  anomalies  to  the  Abelian  Ward  identities, i.e. Equation  (\ref{17}) 
with  $X$   restricted  to  ${\cal A}$, the  Abelian  part of  ${\cal G}$.  Similarly  the consistency
condition  on ${\pmb b}(X) $   leads  to
$${\pmb b}(X)=\mathring \theta(X){\pmb b}$$    
where  ${\pmb b}$     is  arbitrary,  and  can  trivially  be chosen  to  keep the  same  stability group  of its classical limit  (the  residual symmetry group).  Finally  $ {\pmb F}(X) $   is  found  to  be  of
the  form $\mathring{t}(X){\pmb F}$ for  some  fixed  $ {\pmb F}$, up  to terms  which  again can  be  interpreted
as  invariant  anomalies  to  the Abelian  Ward  identities.  

It  will  be
shown,  see   in  particular Appendix  C,  that the Ward identity,  Equation
(\ref{19}),  cannot be  broken  exclusively  by  invariant Abelian  anomalies,  one
concludes  that,  modulo  a  field renormalization,  all  solutions of Equation
(\ref{17}) are  solutions  of  Equation  (\ref{19}).  

We thus  proceed  to  analyze the validity of
Equation  (\ref{19}).  Applying  the  action principle  to the case of the field  variations  given in Equation  (\ref{20}),
we  find
\beq W(X)\Gamma =-\Delta(X) \Gamma\label{21}\eeq 
where  $\Delta(X)$  denotes  the  dimension  four   vertex insertion \footnote{Here $\d_X{\cal L}_{\rm eff}$ is considered  a functional.}
\beq\label{22}  -\Delta (X) =\int dx N_4[ \d_X{\cal L}_{\rm eff} +  \hbar Q(X) ](x)\eeq
where  $\hbar Q(X)$  lumps   the radiative  corrections  together.
Note that the first  one-particle irreducible diagrams appearing in the expansion of  $-\Delta(X) \Gamma$ are the tree diagrams with a single vertex  whose functional generator is \beq-\Delta(X) = \int dx \{ \d_X{\cal L}_{\rm eff} +  \hbar Q(X) \}(x)= W(X)\int dx {\cal L}_{\rm eff}(x) + O(X,\hbar {\cal L}_{\rm eff}),\label{23}\eeq the second term being linear in $X$.
Furthermore, because adding a loop to a diagram introduces a factor $\hbar$, we have the functional equation
\beq\label{23b} -\Delta(X) \Gamma= -\Delta(X) + O(\hbar \Delta(X)).\eeq

The first step of our analysis will consist in deriving  consistency conditions on $\Delta(X)$ 
which stem from the algebraic properties of $W(X)$ (Equation (\ref{20})). Iterating Equation (\ref{21}) we get
\bea \[W(X),W(Y)\]\Gamma &&=W([X,Y]) \Gamma\nn =- \Delta([X,Y]) \Gamma\nn = -\[ W(X)\Delta(Y) \Gamma - W(Y)\Delta(X) \Gamma\],\eea
therefrom, using Equation (\ref{23b}), we get
\beq \label{24} W(X)\Delta(Y)- W(Y)\Delta(X) = \Delta([X,Y]) + \hbar O(\Delta(X),\Delta(Y), \Delta([X,Y])).\eeq
Since  ${\cal L}_{\rm eff}$ and $Q$ belong to  finite-dimensional  representation  spaces  of
$G$, $\Delta$  can be  reduced  into  irreducible  components. 

Equation  (\ref{24})  is  a  perturbed Lie algebra cocycle  condition.\footnote{The following analysis consists in  a perturbed version of the construction of the first class Lie algebra cohomology which is  discussed in Appendix B.} Having  split $ {\cal G} $ into its  Abelian  part  and  its  semi-simple  part:
$ {\cal G} = {\cal A}+{\cal S}$
and  $  \Delta (X)  $  into  its invariant  and  non-invariant  parts:
$$\Delta (X)  = \Delta^\sharp(X)  + \Delta^\flat (X) ,  $$
 let
\beq\label{24bis}  X=X^\a e_\a, \quad W(e_\a)\equiv {\cal T}_\a, \quad \Delta(e_\a)\equiv B_\a \eeq
$ e_\a $  being  a  basis  in  ${\cal G}$.
Due to its linearity in $X$ and $Y$  Equation  (\ref{24})  can  be  rewritten
\beq\label{24t} {\cal T}_\a B_\b- {\cal T}_\b B_\a-f_{\a\b}^\c  B_\c = \hbar{\cal M}_{\a\b} (B),\eeq
Let $\{X,X\}$ be a symmetric, positive definite, invariant form on ${\cal G}$  (e.g. ${\rm Tr}(W(X)W(X))$) which 
can be used to raise and lower indices.
Let  $ \{{\cal T}, {\cal T}\} = {\cal T}_\a {\cal T}^\a $,
we get from Equation  (\ref{24t})  
\beq  \{{\cal T}, {\cal T}\}  B_\b - {\cal T}_\b  {\cal T}^\a B_\a = \hbar {\cal T}^\a{\cal M}_{\a\b} (B) \eeq
where  commutation  relations  have been used  together  with  the antisymmetry  of  $ f_{\a\b\c} $   which  is 
due  to  the  invariance of  $\{X,X\}$. Positive definiteness
of  $\{X,X\}$ insures  that  $\{{\cal T}, {\cal T}\} $   is strictly  positive on the non-invariant ($\flat$)  part,  so  that 
using  again invariance,  which  insures  that
$$ \[{\cal T}_\b, \{{\cal T}, {\cal T}\}\]=0, $$
we  get
\beq  B^\flat_\b = {\cal T}_\b {{\cal T}^\a \o \{{\cal T}, {\cal T}\}^\flat} B_\a^\flat + \hbar  {{\cal T}^\a \o 
\{{\cal T}, {\cal T}\}^\flat}  {\cal M}_{\a\b}^\flat (B)\eeq
i.e.
\beq  \Delta^\flat(X) = W(X)\hat \Delta +O(\hbar \Delta)\eeq
where  $\hat \Delta$  is  linear  in  $\Delta^\flat$.
Furthermore,  for  $e_\a , e_\b \in {\cal S}$
\beq f_{\a\b}^\c B^\sharp_\c =\hbar  {\cal M}_{\a\b}^\sharp (B).\eeq
Thus,  using  the  non-degeneracy  and  invariance of the Killing form for ${\cal S}$, we have
\beq   B^\sharp_\a = O(\hbar B), \quad\quad e_\a \in  {\cal S}.\eeq
Assuming  temporarily  that  also
\beq   B^\sharp_\a = O(\hbar B), \quad\quad e_\a \in  {\cal A},\eeq
as  we shall demonstrate  in  a  moment, we  have
\beq  \label{25}  \Delta(X) =W(X)\hat\Delta+O(\hbar \Delta).\eeq
Now we  show that it is possible to choose ${\cal L}_{\rm eff}$ and $F$ in such a way that
\beq \hat \Delta \equiv {\cal T}^\a B^\flat_\a=0. \label{26bis}\eeq
Indeed Equation (\ref{22}) reads
$$ B_\a = {\cal T}_\a \int dx\ {\cal L}_{\rm eff} + \hbar Q_\a
$$
and  separating in ${\cal L}_{\rm eff}$ and $Q(X)$ the invariant and non-invariant parts, Equation  (\ref{26bis})  reads
\beq \int dx\ {\cal L}_{\rm eff}^\flat +\hbar {{\cal T}_\a \o \{{\cal T}, {\cal T}\}^\flat}  Q^\flat_\a =0,\footnote{Which  one may check to be independent of the choice of $\{X,X\}.$}\eeq
 which is soluble for $ \int dx\ {\cal L}_{\rm eff}^\flat $ in terms of 
 $ \int dx\ {\cal L}_{\rm eff}^\sharp $, ${\pmb F}$ and ${\pmb b}$.\footnote{Note that renormalizabilty implies that $  {\cal L}_{\rm eff}$ depends on a finite number of parameters which are formal power series in $\hbar$ and $Q^\flat$ can be written as a formal power series in ${\cal L}_{\rm eff}$ and $\hbar$.} 

 Once  ${\cal L}_{\rm eff}$ and ${\pmb F}$ are so adjusted, Equation (\ref{24}) is of the form
 $$ \Delta = O(\hbar \Delta) $$
 whose solution is 
 $$\Delta=0. $$ 
The breaking parameter  ${\pmb b}$, which has been so far left  arbitrary,  will  eventually  be determined
together  with  $ {\cal L}_{\rm eff}^\sharp$  in  terms  of  the physical  parameters.

Thus, there remains  to  prove  that
\beq \label {27}\Delta^\sharp =O(\hbar \Delta),\eeq
which  requires a more detailed analysis   than that provided by power counting used  up  to  now. 

The  idea  is  to order  $\Delta^\sharp$  according  to  terms of decreasing  dimensions
and  analyze the  various terms  successively  \cite{ls}\cite{b}.
For  this  purpose,  let  us  consider the linear space spanned by the integrated  monomials
in  the components  of  ${\pmb \varphi}$, ${\pmb \b}$    and  their derivatives. Denoting altogether these   functional variables  by ${ \pmb\Phi}$, we define
\beq\label{28}  {\pmb M}_{I, J, \mu (I\cup J)}\equiv \int dx\prod _{i\in I\cup J}\prod_{\s=0}^3\p_\s^{\mu_\s(i)}\Phi_i (x)\eeq
where $I$ and $J$  denote sets of, possibly repeated, components of ${\pmb \varphi}$ and  ${\pmb \b}$    respectively and  $\mu_\s(i)$ is a
 4-vector  valued function on the union of these sets whose components are integers identifying the degree of the $x^\s$-derivative on the $i$-th element.

It is clear that, on the one hand, different  functions $\mu$  and $\nu$ must be identified if they  coincide after permutations of 
elements of $I$ and $J$ corresponding to the same component of the fields and, on the other hand, that linear combinations of 
${\pmb M}$'s are trivial  if the corresponding linear combinations of the  monomials appearing in 
Equation (\ref{28})  are equal to a total derivative. For this reason we  fix  a unique basis of the space spanned by the monomials by ordering in a given sequence the components of ${\pmb \Phi}$ and we identify one element in the equivalence class  up to a total derivative choosing the monomial in which $\Phi_{i_M}$, the last component of  $\Phi$ belonging to $I\cup J$, appears at least once  without derivatives.

The set of integrated monomials ${\pmb M}$ with  canonical dimension bounded by $d$ (we shall consider in particular
 the case $d=4$) span a finite dimensional linear space in much the same way as polynomials of bounded 
degree are elements of a finite dimensional linear space. The {\it dual space} of the space of polynomials is spanned by multiple
 derivatives at the origin.
In our case we introduce the dual functional differential operators ${\pmb X}$ defined by
\beq {\pmb X}_{I, J, \mu (I\cup J)} ({\pmb q})\equiv {\d \o\d \Phi_{i_M}(0)} \prod _{i\in I\cup J,\  i\not= i_M}\prod_{\s=0}^3\p_\s^{\mu_\s(i)} {\d \o\d\tilde\Phi_i (q_i)},\eeq where $\tilde \Phi$ denotes the Fourier transformed field.
 It  is easy  to  see  that one has the following orthogonality property
\beq \label{C} {\pmb X}_{I, J, \mu (I\cup J)} ({\pmb 0}) {\pmb M}_{I', J', \mu' (I'\cup J')}|_{{\pmb \Phi}=0}= N_{I, J, \mu (I\cup J)}\d_{I,I'} \d_{J,J'} \d_{\mu,\mu'}, \eeq indeed, in particular, the right-hand side of Equation (\ref{C}) vanishes unless $I=I', J=J'$ and hence $i_M=i'_M$. 
$N$ is a non vanishing normalization factor.
Furthermore, 
\beq \label{A} {\pmb X}_{I, J, \mu (I\cup J)} ({\pmb q}) {\pmb M}_{I', J', \mu' (I'\cup J')}|_{{\pmb \Phi}=0}=  {\pmb X}_{I, J, \mu (I\cup J)} ({\pmb 0}) {\pmb M}_{I', J', \mu' (I'\cup J')} |_{{\pmb \Phi}=0}\eeq
  provided that the canonical dimensions
\beq \label{B}  {\rm dim} {\pmb M}_{I, J, \mu (I\cup J)} \geq {\rm dim} {\pmb M}_{I', J', \mu' (I'\cup J')}.\eeq
Let  ${\cal G}$ act  on  ${\pmb X}$  according  to:
\beq {\pmb X}_{I, J, \mu (I\cup J)} ({\pmb q}) {\cal T}_\a^H\bar \Gamma |_{{\pmb \Phi}=0} \equiv \({\cal T}_\a^H {\pmb X}_{I, J, \mu (I\cup J)} ({\pmb q})\)\bar\Gamma|_{{\pmb \Phi}=0}\eeq for any integrated  local functional $\bar \Gamma$.
We have set
\beq  {\cal T}_\a^H\equiv \int dx\Bigr\{ {\pmb\varphi} \mathring{t}^T _\a  {\d \o\d{\pmb \varphi} }
+ {\pmb \b } \mathring{\theta}^T _\a  {\d \o\d{\pmb \b} }\Bigr\}, \eeq and
${\cal T}_\a^H$ is the homogeneous part of ${\cal T}_\a $ obtained
by putting  $ {\pmb \b }={\pmb F }=0$ in  Equation (\ref{20}).
Let 
\beq {\pmb X}^\sharp ({\pmb q}) =\sum_{I, J, \mu (I\cup J)} C^\sharp_{I, J, \mu (I\cup J)}{\pmb X _{I, J, \mu (I\cup J)} }\eeq
be an element of  a basis of ${\cal T}_\a^H$-invariant test operators corresponding to dimension four 
local polynomial functionals, one has
\bea {\pmb X}^\sharp ({\pmb q})  {\cal T}_\a \Gamma|_{{\pmb \Phi}=0}&&= - {\pmb X}^\sharp ({\pmb q})  B_\a \Gamma |_{{\pmb \Phi}=0}\nn
= - {\pmb X}^\sharp ({\pmb q})  \int dx\Bigr\{ {\pmb F} \mathring{t}^T _\a  {\d \o\d{\pmb \varphi} }
+ {\pmb b } \mathring{\theta}^T _\a  {\d \o\d{\pmb \b} }\Bigr\}\Gamma|_{{\pmb \Phi}=0}\nn=
- {\pmb X}^\sharp ({\pmb q})B_\a |_{{\pmb \Phi}=0}  +O(\hbar B)\nn
=- {\pmb X}^\sharp (0) B_\a |_{{\pmb \Phi}=0}  +O(\hbar B).\eea
The  first line  is  a  consequence  of the anomalous  Ward identity,  the  second
one makes use  of  the  ${\cal T}_\a^H$-invariance  of  ${\pmb X}^\sharp ({\pmb q})$, the  last  one follows
from  Equations (\ref{23b}, \ref{A},  \ref{B}).
Now  for  ${\pmb q}$  large  in  the Euclidean region power  counting  insures  that  the
expression
$${\pmb X}^\sharp ({\pmb q})  \int dx\Bigr\{ {\pmb F} \mathring{t}^T _\a  {\d \o\d{\pmb \varphi} }
+ {\pmb b } \mathring{\theta}^T _\a  {\d \o\d{\pmb \b} }\Bigr\}\Gamma |_{{\pmb \Phi}=0}$$ is asymptotically negligible, because ${\rm dim}\ {\pmb\varphi}>0 $  and ${\rm dim}\ {\pmb \b}>0 $, and hence, this expression is a linear combination of multiple derivatives of one-particle irreducible Feynman amplitudes with global dimension, including the field and momentum derivatives, smaller than minus four.  It must vanish for large, linearly independent, ${\pmb q}$'s in the Euclidean region. Thus  
\beq {\pmb X}_{I, J, \mu (I\cup J)} ^\sharp (0) B_\a   =O(\hbar B), \quad \forall (I, J, \mu (I\cup J)),\eeq
and hence the dimension four part of $ B_\a^\sharp  $ is $ O(\hbar B)$.
The analysis of the lower dimension terms of  $ B_\a^\sharp  $ is slightly more sophisticated and is given in Appendix  C.

This analysis completes the proof of Equation (\ref{27})

At  this point,  we have  completed the  construction  of  an effective action
fulfilling  the  Ward identity  (\ref{19}).
The  free  power  series  parameters  $C_\sharp$,  and  ${\pmb b}$  can then  be used  to fulfill
the    normalization  conditions  which  allow  a  particle  interpretation
of  the  theory under the  assumptions  stated  in  section  (2.1), namely  the  existence in the tree approximation of  an  invertible 
 transformation from  $\mathring{C_\sharp}$  and $\mathring{\pmb b}$  to the  physical parameters
 involved in  the  normalization  conditions.

\section{  The local Ward identity ({\pmb{\it  current algebra}})}

\subsection{  The  Classical Theory \cite{cabe}}

Given  a Lagrangian  $\tilde{\cal L}({\pmb\varphi}, {\pmb\b})$  invariant under the global  transformation
Equation  (\ref{20}),  it  is easy  to  introduce  an external  gauge  field $ {\pmb a}_\mu $ of  dimension  1,
and  construct a Lagrangian    ${\cal L}({\pmb\varphi}, {\pmb\b}, {\pmb a}_\mu )$ invariant under  the local  gauge
transformation:
\bea \d_{\pmb\omega} {\pmb \varphi}(x)&&=- \mathring{t} ( {\pmb\omega}(x)) ({\pmb\varphi}(x)+ {\pmb F}),\nonumber\\
\d_{\pmb\omega} {\pmb \b} (x)&&=- \mathring\theta ({\pmb\omega}(x)) ({\pmb\b}(x)+ {\pmb b}),\nonumber\\
\d_{\pmb\omega} {\pmb a}_\mu(x) &&=\p_\mu {\pmb\omega}(x)-[ {\pmb \omega}(x), {\pmb a}_\mu(x)],
\label{29}\eea
 where  we have considered $ {\pmb a}_\mu (x) $  as  well  as ${\pmb\omega}(x)$  as  elements of  ${\cal G}$:  
it  is  enough to  replace  the derivatives  occurring  in  ${\cal L}({\pmb\varphi}, {\pmb\b})$ by  covariant  derivatives:
\bea \p_\mu{\pmb \varphi} \to D_\mu{\pmb \varphi}&&=\p_\mu{\pmb \varphi} + \mathring{t}({\pmb a}_\mu)
({\pmb \varphi}+ \mathring{\pmb F}),\nonumber\\
\p_\mu{\pmb \b} \to \Delta_\mu{\pmb \b}&&=\p_\mu{\pmb \b} + \mathring{\theta}({\pmb a}_\mu)
({\pmb \b}+ \mathring{\pmb b}),
\eea
and  to  include  gauge  invariant  terms constructed  with $ {\pmb a}_\mu$,  through
the antisymmetric  covariant  tensor
$${\pmb G}_{\mu\nu}=\p_\mu {\pmb a}_\nu - \p_\nu {\pmb a}_\mu-[ {\pmb a}_\mu,   {\pmb a}_\nu]. $$
The  local Ward identity  which  expresses  the  invariance  of ${\cal L}({\pmb\varphi}, {\pmb\b}, {\pmb a}_\mu )$
under the local  gauge  transformation  Equation  (\ref{29}) is:
\bea {\cal W} ({\pmb\omega} ) \mathring{\Gamma}({\pmb\varphi}, {\pmb\b}, {\pmb a}_\mu )&&= \int dx \left[{\d \mathring{\Gamma}\o\d{\pmb a}_\mu } (\p_\mu {\pmb\omega} (x)-[ {\pmb \omega}(x), {\pmb a}_\mu(x)])\right.\nn\left.
-{\d \mathring{\Gamma}\o\d{\pmb \varphi} } \mathring{t} ({\pmb\omega}) ({\pmb\varphi}+ \mathring {\pmb F})\right.\nn\left.
-{\d \mathring {\Gamma}\o\d{\pmb \b} } \mathring{\theta} ({\pmb\omega}) ({\pmb\b}+ \mathring {\pmb b})\right]=0,\quad\quad {\pmb\omega}(x) \in{\cal G}.\eea
The  relationship between  the integrated  Ward identity for  ${\cal G}$  and  the  local Ward
identity  for  the  associated gauge  group  is:
$$W( {\pmb \omega} )={\cal W} ({\pmb \omega})\quad {\rm for\ } {\pmb \omega\ } {\rm  space{-}time\  independent}.$$
Note that  the introduction of the  external    gauge  field  ${\pmb a}_\mu $,  which
globally transforms under the  adjoint  representation  of  ${\cal G}$ whose generators are denoted by $\{f_\a\}$,   does
not  spoil    the  conclusions  of  the  previous section  because nowhere  Lorentz  covariance  of  the  fields  was  used.

\subsection{  Radiative  Corrections}

Defining $$ {\cal W}_\a(x) \equiv {\d {\cal W}  \o\d\omega^\a(x)},$$
  we are  going  to give a  general  proof of an anomalous local  Ward identity:
$${\cal W}_\a(x) \Gamma = G_\a (x)$$ 
where $ G_\a$  is  a dimension  four  polynomial  in  the  classical gauge  field ${\pmb a}_\mu$
and  its  derivatives.
Taking into account the remark at the end of  section 2 we have already  proved the integrated  Ward identity  in  the presence  of the  gauge  field:
$$\int dx \Bigr\{ {\pmb a}_\mu  f^T_\a{\d\o\d {\pmb a}_\mu }+ ({\pmb \varphi} + 
{\pmb F} )\mathring{t}^T_\a {\d\o\d {\pmb \varphi} } + ({\pmb \b} + 
{\pmb b} )\mathring{\theta}^T_\a {\d\o\d {\pmb \b} }\Bigr\}(x)\Gamma = 0.$$
On the contrary, performing  a  local  gauge  transformation  yields
$$ {\cal W}_\a(x) \Gamma =   {\cal K}_\a (x) \Gamma$$
where ${\cal K}_\a(x)$   is  a  dimension  four  local insertion.
It  follows  from  the  validity  of  the  integrated  Ward identity that
$$ \int dx\ {\cal K}_\a (x) =0, $$
hence
$$ {\cal K}_\a (x) =\p_\mu {\cal K}^\mu_\a (x)  $$
where  ${\cal K}^\mu_\a$ a  dimension three local operator.
Now the quantum action principle implies that  ${\cal K} _\a$  fulfills the perturbed compatibility condition  \cite{wz}
\beq\label{79}   {\d {\cal K} _\b(y) \o\d\omega^\a(x)} - {\d {\cal K} _\a(x) \o\d\omega^\b(y)} -f_{\a\b}^\c \d(x-y) {\cal K} _\c(x)  =O(\hbar {\cal K} ) \eeq
where  $ O(\hbar {\cal K} )  $  lumps   the radiative  corrections together.

In Appendix D it is shown that the solution to the unperturbed compatibility condition

\beq\label{80}  {\d\hat {\cal K} _\b(y) \o\d\omega^\a(x)} - {\d\hat {\cal K} _\a(x) \o\d\omega^\b(y)} -f_{\a\b}^\c \d(x-y)\hat {\cal K} _\c (x) =0 \eeq   
is of the form
\beq\label{81}\hat {\cal K} _\a = {\d \o\d\omega^\a(x)} \int dy\ K(y) +  G_\a (x) \eeq
where  $ G_\a (x) $ does  not  depend  on  the quantized  fields  and is  not  the  gauge
variation of  any local  functional  of dimension  less  than  or  equal  to  four.  $K(x)$ is a local dimension 
four functional.
Furthermore    the insertion of the vertex  $ G_\a (x) $ into the effective action is additive and does  
not  contribute any  radiative correction\footnote{That is: to the right-hand side of Equation (\ref{79}). Concerning the radiative corrections to $G$ see also  \cite{bar}\cite{cabe} .}
\beq\label{81b} G_\a (x) \Gamma = G_\a (x). \eeq
Therefore,  the  solution  of  Equation  (\ref{79})  is  provided by  
\beq\label{82} {\cal K}_\a(x) = \hat{\cal K}_\a(x) +O(\hbar ({\cal K} - G_\a)).\eeq
Furthermore, since  $ {\cal K} _\a $  is a divergence, so  is $ \hat{\cal K}_\a $.  
  Now we   recall   that, according to the quantum action principle,
\beq\label{83} {\cal K}_\a (x) = {\d \o\d\omega_\a(x)} \int dy {\cal L}_{\rm eff} (y) +  {\cal Q}_\a (x) \eeq
 with $ {\cal Q}_\a=O(\hbar {\cal L}_{\rm eff})$. From this equation, considering Equations (\ref{81}), (\ref{82}) and (\ref{83}), we have
\bea\label{84} {\cal K}_\a (x) &&= {\d \o\d\omega_\a(x)} \int dy\ {\cal L}_{\rm eff} (y) +  {\cal Q}_\a (x)\nn
= \hat{\cal K}_\a(x) +O(\hbar ({\cal K} - G))\nn = {\d \o\d\omega_\a(x)} \int dy \ K(y) +  G_\a (x) + O(\hbar ({\cal K} - 
G_\a)).\eea
From which we have
\bea  {\cal Q}_\a (x) - G_\a (x) &&= {\d \o\d\omega_\a(x)} \int dy  \[K(y)-
{\cal L}_{\rm eff} (y)  \]+ O(\hbar ({\cal K} -  G_\a))\nn\equiv {\d \o\d\omega_\a(x)} \int dy\ N(y)+ O(\hbar ({\cal K} -  G )),\eea where $N(y)$  is  a term generated by radiative corrections and hence is  $O(\hbar {\cal L}_{\rm eff} (y)).$
It follows that the equation
$$  {\d \o\d\omega_\a(x)} \int dy\[ N(y)+ {\cal L}_{\rm eff} (y)\]={\d \o\d\omega_\a(x)} \int dy  K(y)=0 $$ can be solved in terms of the parameters in $ {\cal L}_{\rm eff}$ and hence the system (\ref{84}) reduces to  $$  {\cal K}_\a (x) - G_\a (x) = O(\hbar ({\cal K} -  G )) $$ whose unique solution is
$${\cal K}_\a (x)\Gamma = G_\a (x)\Gamma .$$
At this point, taking into account Equation (\ref{81b}),  we have
$${\cal K}_\a (x)\Gamma  = G_\a (x)\Gamma= G_\a (x),$$
namely we have proved the anomalous Ward identity
$$ {\cal W}_\a (x)\Gamma= G_\a (x). $$
As shown in Appendix D,  $G_\a (x)$ can always be chosen in the form:
\beq\label{85} G_\a (x)=\p_\mu K^\mu_\a (x) \eeq with
\beq\label{86}  K^\mu_\a (x) = \epsilon^{\mu\nu\rho\s}\[D_{\a\b\c}(\p_\nu a_\rho^\b) a_\c^\s + F_{\a\b\c\d}a^\b_\nu a^\c_\rho a^\d_\s\]\eeq
and
\beq\label{87} F_{\a\b\c\d}={1\o12}\[D_{\a\b\eta}f_{\c\d}^\eta + D_{\a\d\eta}f_{\b\c}^\eta + D_{\a\c\eta}f_{\d\b}^\eta \].\eeq
$D_{\a\b\c}$ is a symmetric invariant rank three tensor on ${\cal G}$,  it  parametrizes the general form of the Adler-Bardeen anomaly.
 
 \section{Conclusion} 
 
 We have completed a number of points of Symanzik's program on the renormalization of theories with symmetry breaking.

For  models  without  massless  particles,  we have  been  able  to  deal  with  an
arbitrary  compact  internal  symmetry Lie group,  and  prove  the  integrated
Ward  identities characteristic  of  a  super-renormalizable  breaking with
given covariance. The  corresponding  anomalous  local Ward identity  -  the
functional  expression of  current  algebra  -  is  then proved  in  full  generality
and a compact  formula  exhibited  for  the corresponding  Adler-Bardeen
anomaly.
Our perturbative treatment  fails  if  power  counting  mixes  with  geometry
to  produce e.g. mass rules,  since  in  this  case a  particle  interpretation  of the
theory is  no  longer  possible.  The breakdown  of our  treatment  generated
by this  phenomenon is  quite more dramatic  in  models  involving  massless
particles.  This  happens  in  particular  if,  due  to  tree  approximation  mass
rules there  are  more  massless scalar  fields  than Goldstone  bosons  (pseudo-Goldstone bosons  \cite{wei}).  In  this  case,  even  the construction  of  a  Green
function theory  needs  a  deep  modification  of the  perturbative  scheme  \cite{bbbc}.
Another  limiting  case  which  is  worth  mentioning  occurs  when the
breaking  has dimension  four and, given  the  direction ${\pmb b}$  which  characterizes
the  breaking,      the most  general  invariant  Lagrangian
formed  with  the quantized  field   is not  the most  general  Lagrangian  invariant  under  the
residual symmetry group $H_{\pmb b}$.

\section{Acknowledgements}

The present paper contains a revision of a work published 35 years ago\cite{brs} whose subject was inspired by  K. Symanzik, in particular,  through its exchange of  correspondence  with R. Stora.
For this reason this paper is  dedicated to the memory of both  R. Stora and K. Symanzik.  The author is indebted to his friends A. Blasi, C. Imbimbo, S. Lazzarini and N. Magnoli for  careful readings of different versions of the manuscript.

\appendix 

\newsection{$ \mathring{\pmb F}$ is a covariant function of $ \mathring{\pmb b}$}

A  classical  action  is  viewed as  an integrate local functional  whose  argument  is
indefinitely  differentiable  with  fast  decrease.  In  the  present  case
$$\mathring{\Gamma} ({\pmb \varphi}, {\pmb \b})= \int dx {\cal L} ({\pmb \varphi}+ \mathring{\pmb F}, {\pmb \b}+ \mathring{\pmb b}) (x)
$$ where $ {\cal L}$ is a classical Lagrangian density without 
a  constant  term, i. e. $ {\cal L} ( \mathring{\pmb F},  \mathring{\pmb b})=0
$, defined up to a divergence. We shall limit ourselves to
  renormalizable Lagrangians, according to the conventional power counting theory
through which fields ${\pmb \varphi}$  are  assigned  dimensions connected  with  the  structure  of the  quadratic part of  ${\cal L}$, the  dimension  of  ${\pmb \b}$    being  a  priori  given,
namely Lagrangians  of  positive dimension smaller  than  or  equal  to  four.  If
${\rm dim}{\pmb \varphi }>  0$,  renormalizable  Lagrangians  are  polynomials. Assuming that  ${\cal L}$
has  no term  linear  in  ${\pmb \varphi}$,  
we  see  that  the  integrated  Ward identity,  Equation (\ref{sette}),  is 
only  meaningful  if
\beq\label{a0} {\d\mathring \Gamma\o\d{\pmb \b} } \vert_{{\pmb \varphi}={\pmb \b}={\pmb 0}}\mathring\theta (X)   \mathring{\pmb b}=0 \eeq
which we shall  assume.  The  field  translation  parameter  appropriate  to
get  rid  of  the  term linear  in ${\pmb \varphi}$   from  a  Lagrangian  which is  an  invariant
formed  with  ${\pmb \varphi}+ \mathring{\pmb F}$    and  ${\pmb\b}+ \mathring {\pmb b}$,  of course, depends  on $ \mathring{\pmb b}$ .
  For constant ${\pmb \varphi}$    and vanishing ${\pmb \b}$,   $ {\cal L} ({\pmb \varphi} +  \mathring{\pmb F}, \mathring {\pmb b}) $  is  an invariant polynomial which we denote by  
${\cal F}$. Hence $ \mathring {\pmb F} $ is implicitly defined by
\beq\label{a1} {\p {\cal F} \o\p  {\pmb \varphi} }(  \mathring{\pmb F}, \mathring {\pmb b}  )=0\eeq
and the  Ward identity  implies
\beq\label{a2} {\p {\cal F}  \o\p{\pmb \varphi} } \mathring{t} (X) ({\pmb\varphi} + \mathring {\pmb F} )
+{\p {\cal F}  \o\p  \mathring{\pmb b}  } \mathring\theta (X) \mathring {\pmb b} =0. \eeq Differentiating  Equation  (\ref{a1})  with  respect  to $ \mathring {\pmb b}  $  and  Equation  (\ref{a2})  
with  respect  to  ${\pmb\varphi} $  at  ${\pmb\varphi} = \mathring {\pmb F}   $ yields
$$ {\p^2 {\cal F}\o (\p {\pmb\varphi})^2}\vert_{{\pmb\varphi} = \mathring {\pmb F} } \[ \mathring{t} (X) \mathring {\pmb F} -{\p \mathring {\pmb F} \o \p \mathring {\pmb b} } \mathring\theta (X) \mathring {\pmb b} \]=0 $$
which, under  the  assumption  that  the  mass  matrix ${\p^2 {\cal F}\o (\p {\pmb\varphi})^2}\vert_{{\pmb\varphi} = \mathring {\pmb F} } $
be non-degenerate,  implies  that  $ \mathring{\pmb F} $ is  a  covariant  function of $ \mathring {\pmb b}$ :
$$  \mathring{t} (X) {\pmb F} = {\p {\pmb F} \o \p {\pmb b} } \mathring\theta (X) {\pmb b}. $$
Similarly the other coefficients of ${\cal L}$ are covariant functions of $ \mathring{\pmb b}$.

\newsection{Canonical form of the Ward Identity}
This appendix  is  devoted to  the  reduction  of the  Ward identity  to  canonical
form.
\subsection{The tree approximation}

We have assumed that ${\cal G } ={\cal S }+{\cal A} $ be a compact Lie algebra and hence that 
$\{X,X\}$ be a symmetric, positive definite, invariant form on ${\cal G}$ which can be used to raise and lower indices.  
We first show that the representations
\beq X\to  \mathring t (X) , \quad X\to  \mathring\theta(X), \quad X\in {\cal G },\eeq
are  fully  reducible. This is automatic  for  $X\in {\cal S} $,  the  semi-simple  part  of  {\cal G }.
For $X\in {\cal A} $,  the  Abelian  part  of  {\cal G }, this  is  a consequence  of the assumption
that  the  kinetic part  of  ${\cal L }$  be  Hermitian  non-degenerate,  which  insures  that
$ X\to  \mathring t (X) $  is  fully  reducible. Then, the Lie algebra cocycle  condition Equation (\ref{13} c) can be solved  as  follows.
Reducing  Equation (\ref{13} c)  to  components (see Equation (\ref{24bis})):
$$  \mathring t_\a \mathring {\pmb F}_\b - \mathring t_\b \mathring {\pmb F}_\a - f_{\a\b}^\c \mathring {\pmb F}_\c = 0 $$
yields
$$ \{ \mathring t, \mathring t\}   \mathring{\pmb F}_\b\equiv  ( \mathring t_\a  \mathring t^\a) \mathring {\pmb F}_\b = \mathring t_\b ( \mathring t^\a \mathring {\pmb F}_\a) . $$
Thus  restricting  $ \mathring {\pmb F}_\a $ to its non-invariant  part  $ \mathring {\pmb F}_\a^\flat $we have:
$$ \mathring {\pmb F}_\a^\flat = {1\o \{ \mathring t, \mathring t\}  }  \mathring t_\a ( \mathring t^\b  \mathring {\pmb F}_\b)\equiv  \mathring t_\a \mathring {\pmb F}^\flat .$$
Similarly,  using the  non-degeneracy of  the  Killing  form of  ${\cal S}$,    
  we also get   $$  \mathring{\pmb F}_\a^\sharp=0, \quad\quad e_\a\in {\cal S}$$
so  that  only $ \mathring{\pmb F}_\a^\sharp $ for $e_\a \in {\cal A}$  is left  undetermined.

Thus one has    to  find  a polynomial Lagrangian  ${\cal L}$  invariant under
$$\d_X {\pmb \varphi}^\flat= - \mathring t(X) ({\pmb \varphi}^\flat + \mathring {\pmb F}^\flat),\quad
\d_X {\pmb \varphi}^\sharp  = \mathring {\pmb F}^\sharp (X) ,\quad
\d_X {\pmb \b}^\flat= -  \mathring\theta(X) ({\pmb \b}^\flat + \mathring {\pmb b}^\flat) . $$
According to the mathematical meaning of  ${\pmb \b}$, which characterizes the symmetry breaking,  any 
${\pmb \b}^\sharp$ component is excluded. Thus the last of the above equations  is proved in the same way as the first one.

It is easy to see that, due to the polynomial character of  ${\cal L}$, those components of  ${\pmb \varphi}^\sharp  $ for which $ \mathring{\pmb F}^\sharp (X)\not= 0$ {\it do not couple}.

\subsection{  Radiative Corrections}
We  shall  first  show  that  the  representation  property Equation (\ref{13}a) \footnote{We replace into Equation (\ref{13}a) $\mathring T(X)$ by $t(X)$ because we are now considering radiative corrections and hence formal power series in $\hbar$. }
with
$$ t(X) = \mathring{t}(X) + O(\hbar)$$
where  $ t(X) $   is  a  formal  power  series  in  $\hbar$, implies that
$$ t(X) =Z^{-1}  \mathring{t}(X) Z$$
for  some  formal power  series  $Z$:
$$ Z  = \mathbb{1} +  O(\hbar) . $$
Let first $X \in {\cal S} $ the  semi-simple  part  of  ${\cal G} $, let
$$ t(X) = \sum_{n=0}^\infty  t_n( X)  $$
$$ Z = \sum_{n=0}^\infty  Z_n   $$
be  the  formal  power series for $t(X)$  and $  Z$,  respectively.  We have chosen
$$ t_0(X)\equiv  \mathring t(X), \quad Z_0  =\mathbb{1}$$
thanks  to  a symmetric  wave  function  renormalization.  The   possibly non-trivial first  order term   in
the  expansion  of Equation (\ref{13}a) reads
$$ [\mathring{t}(X) , t_1 (Y)] - [\mathring{t}(Y) , t_1 (X)] - t_1([X,Y]) = 0 $$ which is a Lie algebra cocycle condition strictly analogous to  Equation (\ref{13}c)\footnote{It just refers to a different representation, the adjoint,  of $ {\cal G} $.} and can be solved in the same way; hence, due to the semi-simplicity of  $ {\cal G} $
$$ t_1 (X)  = [ \mathring{t}(X) , Z_1 ] $$ for some $Z_1$.

Let us now assume that 
$$ t(X) =( Z^{(n-1)} )^{-1}\mathring{t}(X) Z^{(n-1)} +\tau_n (X) $$
 with  $ \tau_n (X) =O(\hbar^n) $  and $ Z^{(n-1)}=\sum_0^{n-1} Z_k $ which is true for $n=2  $ with 
$$Z^{(1)} = \mathbb{1} + Z_1.$$
The term in Equation (\ref{13}a) at the lowest non-vanishing order reads:
\bea&&  \[\mathring{t}(X), \{ Z^{(n-1)} \tau_n (Y) ( Z^{(n-1)} )^{-1}\}_n\]\nn - 
\[\mathring{t}(Y),  \{ Z^{(n-1)}\tau_n (X) ( Z^{(n-1)} )^{-1}\}_n\]\nn -
\{ Z^{(n-1)} \tau_n ([X,Y]) ( Z^{(n-1)} )^{-1}\}_n = 0, \eea
where, given a $\hbar$ formal power series $X$, $\{X\}_n$ denotes the term of order $n$.
This is a further cocycle condition whose solution is
$$\{ Z^{(n-1)} \tau_n (X) ( Z^{(n-1)} )^{-1}\}_n = 
[\mathring{t}(X), Z_n] $$
 for some $Z_n$, so that
$$ \tau_n (X) =( Z^{(n-1)} )^{-1}[\mathring{t}(X), Z_n]  Z^{(n-1)} + O(\hbar^{n+1})=[\mathring{t}(X), Z_n]   + O'(\hbar^{n+1})$$ 
thus
$$ t_n(X) = \{( Z^{(n )} )^{-1}\mathring{t}(X) Z^{(n )} \}_n $$ with $$ Z^{(n )} = Z^{(n-1 )} + Z_n . $$
As  a  conclusion,  we  may  choose  $ t(X) = \mathring{t}(X)  $  for $ X\in {\cal S}$  up  to  a  field  
renormalization identified by $Z$.

Now, for  $ X\in {\cal A}  $,  the  Abelian  part of  ${\cal G}$,  any anomaly  in  the Ward
identity  can  be considered  as  a  breaking  of the canonical
Ward identity
through  a term which,  up  to   $O(\Delta^2)$,  is Abelian
invariant,  and  thus,
cannot occur  as  a  consequence  of  the argument  at  the end  of  section  2
for  the anomaly  in $t(X)$,  and  of  the argument  in next Appendix  C  for  the anomaly in ${\pmb F}(X)$.

\newsection{ Elimination  of Soft  Invariant  Anomalies  from  the  Integrated  Ward  Identity
}

Once the  dimension  four  anomalies have  been  eliminated  as  indicated in
the text, one  might remain with  a Ward  identity  of the form:$$ W(X ) \Gamma = \sum_{\d=1,2,3} \Delta^\sharp_\d(X) \Gamma\quad {\rm for}\quad\  X\in {\cal A}$$
where  the  breaking insertions  $\Delta^\sharp_\d$  are  invariant  and have power counting dimension (Zimmermann's index) $\d$.   
Let now $ \lambda $ be  any  parameter of the  theory (every parameter identifies an independent term of ${\cal L}$),  and  let
\beq\label{c3} D_\lambda= \p_\lambda - \int dx\  \p_\lambda {\cal B} {\d \o \d\Phi (x)}\eeq
where
$$ {\cal B} = ( {\pmb F} , {\pmb b}), \quad\Phi = ( {\pmb \varphi} , {\pmb \b}). $$
Then
$$ \[ D_\lambda , W(X) \] = 0. $$
In  particular, let  $ m \p_m$ be  the  operator which  scales  all  the parameters  of  the
theory  according  to  their  mass dimensions  (the  first  term  in  the  Callan-Symanzik equation) and  $ D_m $  the associated  invariant  operator \cite{b}  (as in  Equation  (\ref{c3})). In  the  tree approximation  $ D_m {\cal L} $ is  invariant and  soft.  

A differential  scaling equation is written introducing   into  ${\cal L}$  an invariant  external fields  $\eta$, with  dimension  $d=1$,  coupled  to soft invariant terms constrained by the condition   for the classical action $\mathring\Gamma$ 
$$ D_m\mathring\Gamma(\eta) = \int dx\ m {\d \mathring\Gamma(\eta) \o\d\eta (x)},$$ where $m$  defines a reference mass scale. 
If this equation is satisfied ${\cal L}$ is a linear combination of  dimension four  independent invariant local  polynomials in  $\Phi$ and in $\eta-m$. The coefficients of this linear combination, that we label by $\xi$, are dimensionless and are constrained by a sum rule which follows from the already stated condition that ${\cal L}$ must vanish  when all the quantized and external fields vanish.\footnote{It is important to  note here that this condition, holding true  in the tree approximation, remains fulfilled also by the loop corrections since the subtraction prescription does not contribute any constant term. }

After the introduction of $\eta$, repeating the analysis  shown in  the  text,  we see that  the    Ward identity becomes
\beq\label{c1} W(X) \Gamma(\eta) = \sum_{\d=1,2,3} \Delta^\sharp_\d(X,\eta) \Gamma (\eta) \eeq
where $ \Delta^\sharp_\d(X,\eta)$ are  the new, soft,  $\eta$-dependent,  invariant  breaking insertions  and
  $\Gamma (\eta)$  the  new $\eta$-dependent  effective action   functional.
  
Considering the scaling equation beyond the tree approximation, we deduce from the quantum action principle\footnote{Which in the present case 
corresponds to the Zimmermann identities giving the expansion of local operators 
with a weaker subtraction prescription  in terms of local operators with stronger subtraction prescriptions, such as those coupled to   $\eta-m$. This difference vanishes in the tree approximation  because there is no diagram to subtract. }
\beq\label{c7}  \[D_m -\int dx\ m {\d \o\d\eta (x)}\]\Gamma (\eta)=
\hbar\[M_4^{m \sharp} (\eta) \Gamma (\eta)   + M_4^{m \flat} (\eta) \Gamma (\eta) \] \eeq
where $ M_4^{m \sharp/\flat} \Gamma(\eta)=\int dx N_4[M^{m \sharp/\flat}(x)]  \Gamma(\eta)$ correspond to the insertion
 into $\Gamma$ of a linear combination of, invariant/non-invariant, integrated local vertices among which there are some which are   $\eta$ dependent. 
 
Furthermore  we have
\beq\label{c7b} D_{\xi}\Gamma (\eta) =M_4^{\xi \sharp} (\eta) \Gamma (\eta)   + \hbar M_4^{\xi \flat} (\eta) \Gamma (\eta),\eeq where the non-invariant ($\flat$) operators appear because the Ward identity is broken.
 
The mentioned operator set (i.e. that spanned by the linear combinations of the $M_4^{\xi \sharp} (\eta)$'s) being complete, there must be a linear relation among  $ M_4^{m \sharp}$ and the $ M_4^{\xi \sharp}$'s. Thus Equation (\ref{c7}) reads
\beq\label{c8}  \[D_m - \int dx\ m {\d \o\d\eta (x)}+\hbar \sum c_{\xi} D_{\xi}  \]\Gamma (\eta)= 
\hbar \tilde M_4^{m \flat} (\eta) \Gamma (\eta), \eeq where $\tilde M_4^{m \flat}$ lumps the non invariant insertions appearing in the right-hand side of Equations (\ref{c7}) and (\ref{c7b}) together.

It is obvious that, if the Ward identity were unbroken, the right-hand side of 
Equation (\ref{c8}) would vanish because it is not invariant, while the left-hand side is. In that case Equation (\ref{c8}) coincides with the Callan-Symanzik equation of the theory (\cite{b}). If, on the contrary, we have the broken Equation (\ref{c1}), combining this equation with Equation (\ref{c8}) we get
\bea \label{c9}&&W(X)  \[D_m- \int dx \ m {\d \o\d\eta (x)}+\hbar \sum c_{\xi} D_{\xi} \]\Gamma (\eta)\nn = \[D_m -  \int dx \ m {\d \o\d\eta (x)}\]\sum_{\d=1,2,3} \Delta^\sharp_\d(X,\eta) + O(\hbar  \Delta^\sharp)\nn =
-\hbar T(X) \tilde M_4^{m \flat} (\eta) + O(\hbar^2\tilde M_4^{m \flat}). \eea Indeed the last term in the second line is due to the action of $\hbar\sum c_{\xi} D_{\xi}$  
on the breaking $\Delta^\sharp(\eta)$  and to the loop diagrams with the insertion of $\Delta^\sharp$.  
The last line in Equation (\ref{c9})   accounts for the action of $W(X)$ on the right-hand side of Equation (\ref{c8}) whose first order approximation is given by the action of $T(X)$ on $\hbar  \tilde M_4^{m \flat} (\eta)$.

Equation (\ref{c9}) is equivalent to a system of equations involving  terms with different dimensions and covariances. For the non-invariant part we have
$$ T(X) \tilde  M_4^{m  \flat} (\eta) \sim \tilde M_4^{ m \flat} (\eta) = O(\Delta^\sharp,  \hbar M_4^{m \flat} (\eta)) $$ which implies 
$$\tilde M_4^{m \flat} (\eta) = O(\Delta^\sharp  ) .$$ Thus we have
$$ \[D_m -  \int dx \ m {\d \o\d\eta (x)}\]\sum_{\d=1,2,3} \Delta^\sharp_\d(X,\eta) = O(\hbar  \Delta^\sharp).$$
Now, considering in the order the terms with decreasing powers of $\eta$ and decreasing dimension $d$, none of which is annihilated by the differential operator $D_m$\footnote{Indeed the breaking $\Delta_\d$ has physical (mass) dimension 4 and power counting dimension $\d\leq 3$, this implies the presence of coefficient with mass dimension larger than one.}, we finally get $$ \Delta^\sharp_\d(X,\eta)\equiv 0$$ and the unbroken integrated Ward identity is proved.

\section {Cohomology  of  the  Gauge  Lie  Algebra }

We  shall  analyze the structure  of
\beq\label{d1} \hat{\cal K} _\a(x) = \p_\mu K_\a^\mu \eeq
solution  of  the gauge algebra\footnote{The gauge Lie algebra discussed in the present paper is an infinite dimensional generalization of a Lie algebra. The analysis shown in this section has been extended to a more general situation in \cite{bbh}. } cocycle  condition  (Cf.  Equation  (80))
\beq\label{d2} {\d\hat {\cal K} _\b(y) \o\d\omega_\a(x)} - 
{\d\hat {\cal K} _\a(x) \o\d\omega_\b(y)} -f_{\a\b}^\c \d(x-y)\hat {\cal K} _\c (x) =0 \eeq 
(the  Wess-Zumino  consistency  condition  \cite{wz}).
Integrating first  Equation (\ref{d2}) over $x$ shows that $ \p_\mu K_\a^\mu$  transforms
like  the  adjoint (regular) representation, under global transformations.  Indeed,
upon $x$-integration,  Equation (D.2) reduces to 
\beq\label{d3} T_\a \hat{\cal K} _\b(y) - f_{\a\b}^\c \hat{\cal K} _\c(y) \eeq
where $ T_\a$ is the infinitesimal generator of the global transformations. 
Due to its definition, Equation (\ref{d1}), $K ^\mu_\a$ is a local polynomial in the fields and their derivatives identified up to terms which belong to the kernel of $\p_\mu.$  The mentioned polynomials carry completely reducible representations of the compact Lie algebra ${\cal G}$ which commutes with $\p_\mu.$  Thus, writing  $K_\a^\mu$ as a combination of terms, each belonging to a different irreducible representation ${\cal G}$, we should find two different combinations corresponding to the kernel of $\p_\mu $ and to the rest of  $K_\a^\mu$  which must belong to the adjoint representation. In the following we shall only consider this rest which we shall persist denoting by   $K_\a^\mu$ and which belongs to the adjoint representation of ${\cal G}$
.
We shall  now  expand $K^\mu_\a$ in increasing  powers of ${\pmb a}_\mu$, obviously every term of this expansion
belongs to  the adjoint representation of ${\cal G}$. 
Let   {\fns $ K^\mu_{0\a} $} be the term independent of ${\pmb a}_\mu$. We may write
\beq\label{d10b}\p_\mu   K^\mu_{0\a} = {\d\o\d \omega^\a (x)} \int dy \  K_{0\b}^\nu(y)a_\nu^\b(y)
 + \p_\mu L^\mu_\a\eeq where $ L^\mu_\a $ subtracts the homogeneous part of the gauge transformation of the first term and hence  is linear in ${\pmb a}_\mu$.
The same decomposition can be repeated for the terms of higher degree.

Let   $  K_{1}^\mu $ be the term of $ K^\mu  + L^\mu $ linear in ${\pmb a}_\mu$. We can similarly write
$$ \p_\mu K^\mu_{1\a} (x) ={1\o2} {\d\o\d \omega^\a (x)} \int dy \  K_{1\b}^\nu(y)a_\nu^\b(y)
 + \p_\mu Q^\mu_\a$$
provided that
\beq\label{d11}  K^{\mu\nu}_{1\a\b} (x,y) \equiv {\d  K^\nu_{1\b} (y) \o\d a_\mu^\a (x)} - 
{\d  K^\mu_{1\a} (x) \o\d a_\nu^\b (y)} = 0. \eeq
$ Q^\mu_\a $ is now quadratic in ${\pmb a}_\mu$.
From  the  ${\pmb a}_\mu$ independent part of  Equation  (\ref{d2})  we get
\beq\label{d12} \p_\mu^x\p_\nu^y K^{\mu\nu}_{1\a\b} (x,y) = 0. \eeq
The only possible   $K^{\mu\nu}_{1\a\b} (x,y)$ consistent with power counting, symmetry, and condition (\ref{d12}), is:
$$K^{\mu\nu}_{1\a\b} (x,y) = (\square g^{\mu\nu}-\p^\mu\p^\nu )\d(x-y) A_{\underline{\a\b}} $$
for some $A_{\underline{\a\b}}$  anti-symmetric invariant tensor on the Lie algebra. Then
$$  {\tilde K}^{\mu}_{1\a} (x) =  K^\mu_{1\a} (x) + (\square a^{\mu}_{\b}(x)-\p^\mu\p_\nu a^{\nu}_{\b} (x)) A_{\underline{\a\b}} $$ does fulfill Equation (\ref{d11}) and 
$$\p_\mu {\tilde K}^{\mu}_{1\a} (x) = \p_\mu 
K^{\mu}_{1\a} (x).$$ Thus
\beq\label{d16b} \p_\mu K^\mu_{1\a} (x) ={1\o2} {\d\o\d \omega^\a (x)} \int dy \ {\tilde K}_{1\b}^\nu(y)a_\nu^\b(y)
 + \p_\mu R^\mu_\a(x)\eeq for some $ R^\mu_\a(x) $  quadratic in ${\pmb a}_\mu$.
 
 Similarly we proceed considering the terms {\fns $K ^{\mu}_{2\a} (x) =K ^{\mu}_{\a} (x)+ R ^{\mu}_{\a} (x) $} quadratic in  ${\pmb a}_\mu$.
 
 It is convenient to continue our analysis after Fourier transformation of fields and local functionals. To simplify our formulae and calculations for a generic quantity $f(x)$ (or $y$) we denote  its Fourier transform by $f(p)$ (or $q$ or else $k$), only  changing   the variables.
 The most general form of {\fns $ K ^{\mu}_{2\a} (-p)$} which is not orthogonal to $p$ is
\bea\label{d17}&&K ^{\mu}_{2\a} (-p)=i \int dk \[k^\mu a^\nu_\b (k)  a_{\nu \c}(-p-k)) Z^{\a\b\c} + k^\nu a^\mu_\b (k)  a_{\nu \c}(-p-k)) X^{\a\b\c}\right.\nn\left. +  k^\nu a_{\nu\b} (k)  a^\mu_{ \c}(-p-k)) Y^{\a\b\c} + 
 \epsilon^{\mu\nu\rho\s} D^{\a\b\c}  k_\nu a_{\rho \b} (k) a_{\s\c}(-p-k)\],  \eea where   $D^{\a\b\c}$  must be symmetric in $\b$ and  $ \c$ \footnote{\label{8}In order to verify these properties it is useful to have occasionally recourse  to the  change of the integration variable $k\to-p-k$.}.
 Furthermore all the coefficient  are invariant tensors on ${\cal G}$.

 Now
 \bea\label{d18}&&ip_\mu K ^{\mu}_{2\a} (-p)= - \int dk \[p\cdot k a^\nu_\b (k)  a_{\nu \c}(-p-k)) Z^{\a\b\c} +p_\mu k^\nu a^\mu_\b (k)  a_{\nu \c}(-p-k)) X^{\a\b\c}\right.\nn\left. +  p_\mu  k^\nu a_{\nu\b} (k)  a^\mu_{ \c}(-p-k)) Y^{\a\b\c} + 
 \epsilon^{\mu\nu\rho\s} D^{\a\b\c}  k_\mu a_{\nu \b} (k) p_\rho a_{\s\c}(-p-k)\],  \eea 
    The part of the cocycle (consistency) condition on {\fns $p_\mu K ^{\mu}_{2\a}(-p) $} which   is linear in ${\pmb a}_\mu$ requires the symmetry under simultaneous interchange of $p$ and $q$ and $\a$ and $\b$ of 
\bea\label{d19}  q_\nu {\d\   p_\mu K ^{\mu \a}_2 (-p)\o\d\  a_{\nu\b}(q) }&&=q\cdot a_{\c}(-p-q)[p\cdot q ( Z^{\a\b\c}- Z^{\a\c\b}- Y^{\a\c\b}+X^{\a\b\c})-p^2 Z^{\a\c\b}]\nn
+p\cdot a_{\c}(-p-q)[p\cdot q ( - Y^{\a\c\b}-X^{\a\c\b})-q^2( X^{\a\c\b}- Y^{\a\b\c})]
\eea where we have  performed twice the  partial change of variables mentioned in the footnote.
  From Equation (\ref{d19}) we get
  \bea \label{d20}&& Z^{\a\c\b}=X^{\b\c\a}-Y^{\b\a\c}\nn Z^{\a\b\c}- Z^{\a\c\b}- Y^{\a\c\b}+X^{\a\b\c}+Y^{\b\c\a}+X^{\b\c\a}=0.
   \eea

 We now consider {\footnotesize $K_3$}, the most general integrated local functional of dimension four and cubic in ${\pmb a}_\mu$, it is
  \bea K_3&&=\int dx\ \[\p^\mu a_{\mu\b}(x)a^{\nu}_{\c}(x)a_{\nu\d}(x) A^{\b\c\d}+ a_{\mu\b}(x)\p^\mu a^{\nu}_{\c}(x)a_{\nu\d}(x) B^{\b\c\d}\right.\nn\left.+
   \epsilon^{\mu\nu\rho\s} E^{\b\c\d} \p_\mu a_{\nu\b}(x)a_{\rho\c}(x)a_{\s\d}(x)\]
  \eea where $A^{\b\c\d}$ is symmetric in $\c$ and $\d$ and $E^{\b\c\d} $ is anti-symmetric in the same indices. Computing $p_\mu {\d\    K_3\o\d\  a_{\mu\a}(p) }$ we get the same expression as that in the right-hand side of Equation (\ref{d17}) where however 
  \beq \label{d21} Z^{\a\b\c}=B^{\a\b\c}-2 A^{\a\b\c}=-Y^{\b\a\c}, \quad {\rm and}\quad X^{\a\b\c}=Z^{\c\b\a}-Z^{\c \a\b}. \eeq These are consistent with Equation (\ref{d20}).
 $B$ being arbitrary, although invariant,  we can choose $B$ and $A$  satisfying Equation (\ref{d21}). With this choice and using Equation (\ref{d18}) we have, for some $Q_3$ cubic in ${\pmb a}_\mu$
 \bea\label{d22}&&ip_\mu K ^{\mu\a}_{2} (-p)={\d\o\d \omega_\a (p)}  K_3 + Q^\a_{3}(p)\nn
 -\int dk\ p_\mu k^\nu \[ a^\mu_\b (k)  a_{\nu \c}(-p-k)) (X^{\a\b\c}-Z^{\c\b\a}+Z^{\c \a\b}) \right.\nn\left. +   a_{\nu\b} (k)  a^\mu_{ \c}(-p-k)) (Y^{\a\b\c} +Z^{\b\a\c})\]\nn- 
 \int dk\ \epsilon^{\mu\nu\rho\s} (D^{\a\b\c}-2(E^{\a\b\c}+E^{\b\a\c} )) k_\mu a_{\nu \b} (k) p_\rho a_{\s\c}(-p-k)\eea
 Using the first Equation (\ref{d19}) we get $X^{\a\b\c}-Z^{\c\b\a}+Z^{\c \a\b}=Y^{\a\c\b} +Z^{\c\a\b}\equiv W^{\a\c\b}$ which is antisymmetric in the last two indices due to   Equation  (\ref{d19}). Now it is not difficult to verify, using again the above mentioned change of variables, that the third term in the right-hand side of Equation (\ref{d22}) vanishes.
 We are still free to choose $E$; we set $ E^{\a\b\c} = {1\o3}[D^{\b\c\a}-D^{\c\a\b}]$ ($D$ is symmetric in the last two indices). Then the coefficient
  in the  third term in the right-hand side of Equation (\ref{d22}) reads
  $${1\o3}(D^{\a\b\c}+4D^{\c\a\b})-{2\o3} D^{\b\c\a}.$$ However we must remind that the non-vanishing contribution to Equation (\ref{d22}) of its fourth term  corresponds to  the part of this tensor which  is $\b-\c$-symmetric, that is
  \beq \label {d23} \tilde D^{\a\b\c}\equiv {1\o3}(D^{\a\b\c}+D^{\c\a\b}+ D^{\b\c\a}),\eeq
  this is an invariant fully symmetric tensor on the Lie algebra ${\cal G}$.
Thus we get
 \beq\label{d23} \p_\mu K ^{\mu\a}_{2} (x)={\d\o\d \omega_\a (x)}   K_3 + Q^\a_{3}(x)+ 
 \epsilon^{\mu\nu\rho\s} \tilde D^{\a\b\c} \p_\mu a_{\nu \b} (x) \p_\rho a_{\s\c}(x).\eeq
 In order to perform the last step we put together Equations (\ref{d10b}), (\ref{d16b}) and (\ref{d23}) and, omitting the $\tilde{}$ sign above $D$, we obtain
 \bea\label{d24}&& \p_\mu K ^{\mu\a} (x)={\d\o\d \omega_\a (x)}   \[ \int dy \ ( K_{0\b}^\nu(y)a_\nu^\b(y)+{1\o2}  \ {\tilde K}_{1\b}^\nu(y)a_\nu^\b(y))+K_3\]\nn +\p_\mu S^{\mu\a}(x)+ 
 \epsilon^{\mu\nu\rho\s}  D^{\a\b\c} \p_\mu a_{\nu \b} (x) \p_\rho a_{\s\c}(x) = {\d\o\d \omega_\a (x)}   M +\p _\mu J^{\mu\a}(x), 
 \eea
   for some  $S^{\mu\a}$ and hence $J^{\mu\a}$ of dimension four and cubic in ${\pmb a}_\mu$.  Thus   
   \bea\label{d25} \p _\mu J^\mu_\a(x)&&=\p_\mu \[ \epsilon^{\mu\nu\rho\s}  ( {1\o2} D^{\a\b\c} a_{\nu \b} (x) \overset{\leftrightarrow} \p_\rho a_{\s\c}(x)+F^{\a\b\c\d}  a_{\nu \b} (x) a_{\rho \c} (x) a_{\s\d}(x))\right.\nn\left. +G^{\a\b\c\d}  a_{\mu \b} (x) a_{\nu \c} (x) a_{\d}^\nu(x)\],\eea for some $F^{\a\b\c\d}$ and $G^{\a\b\c\d}$  invariant tensors on the Lie algebra ${\cal G}$. $F^{\a\b\c\d}$ is antisymmetric in its last three indices while $G^{\a\b\c\d}$ is symmetric in its last two indices.
 Furthermore  $\p _\mu J^{\mu\a}(x)$ must satisfy the consistency condition (\ref{d2}). 
 
 This condition generates a system of algebraic equations for the coefficients $F^{\a\b\c\d}$ and $G^{\a\b\c\d}$. In particular, the parts containing the antisymmetric four dimensional Ricci symbol give three independent equations that we now write in terms of  space-time functionals
\beq\label{d26}\  \epsilon^{\mu\nu\rho\s} \p_\mu \d(x-y) \p_\nu( a_{\rho \c}  a_{\s \d}) (y)\[3(F^{\a\b\c\d} +F^{\b\a\c\d} )+D^{\a\u\c}f^{\b\d}_\u +D^{\b\u\c}f^{\a\d}_\u\]=0, \eeq
\beq\label{d27} \epsilon^{\mu\nu\rho\s}\p_\mu\d(x-y)( a_{\rho \c} \overset{\leftrightarrow} \p_\nu a_{\s \d}) (y)\[D^{\a\u\c}f^{\b\d}_\u -{1\o2}D^{\d\u\c}f^{\a\b}_\u\]=0, \eeq
 \beq\label{d28} \epsilon^{\mu\nu\rho\s}\p_\mu\d(x-y) (a_{\nu\c} a_{\rho \d}  a_{\s \eta}) (y)\[3f^{\b\d}_\u F^{\a\u\c\eta} -f^{\a\b}_\u F^{\u\d\c\eta}\]=0. \eeq
The equations for the coefficients in Equations (\ref{d27}) and  (\ref{d28}) are
$${1\o2}\[D^{\a\u\c}f^{\b\d}_\u +D^{\a\u\d}f^{\b\c} +D^{\d\u\c}f^{\b\a}_\u\]=0$$  and $$ f^{\b\d}_\u F^{\a\u\c\eta} +f^{\b\c}_\u F^{\a\d\u\eta}+f^{\b\eta}_\u F^{\a\d\c\u}+f^{\b\a}_\u F^{\u\d\c\eta}=0$$ which are trivially satisfied due to the invariance  of $D$ and $F$.

The equation for the coefficient in Equation (\ref{d26}) is
\bea &&\label{d29} 6(F^{\a\b\c\d} +F^{\b\a\c\d} )=D^{\a\u\c}f^{\d\b}_\u +D^{\b\u\c}f^{\d\a}_\u+D^{\a\u\d}f^{\b\c}_\u +D^{\b\u\d}f^{\a\c}_\u\nn
={1\o2}\[D^{\a\u\c}f^{\d\b}_\u +D^{\a\u\d}f^{\b\c}_\u +D^{\b\u\d}f^{\a\c}_\u+D^{\b\u\c}f^{\d\a}_\u+2D^{\b\u\a}f^{\c\d}_\u\],\eea
which is apparently solved by Equation (\ref{87}).

  It is clear that this is a particular solution of Equation (\ref{d29}) whose general solution is obtained by adding to the right-hand side of Equation (\ref{87}) a solution of the corresponding homogeneous equation ($D=0$). This must be  antisymmetric in $\a - \b$. But  $F^{\a\b\c\d}$ is also completely antisymmetric in its last three indices, thus the solution of the homogeneous equation  must be completely antisymmetric in all its indices. However the contribution to $\p\cdot K^\a(x)$ corresponding to  a generic invariant totally antisymmetric $F_A^{\a\b\c\d}$  is just equal to
$${1\o4}{\d\o\d \u_\a(x)}\int dy \epsilon^{\mu\nu\rho\s} (a_{\mu\a}a_{\nu\b} a_{\rho \c}  a_{\s \d}) (x)F_A^{\a\b\c\d} \equiv {\d\o\d \u_\a(x)} N.$$  $N$ can be added to $M$ in Equation (\ref{d24}).

Still we have to discuss the last term in Equation (\ref{d25}), that is the consistency condition (\ref{d2}) for $K_G^\a(x)= G^{\a\b\c\d}  \p^\mu(a_{\mu \b} (x) a_{\nu \c} (x) a_{\d}^\nu(x))$. We find, once again  a system of algebraic equations for the coefficient $G^{\a\b\c\d}$. Selecting the independent parts we have 
\beq\label{d30} \p^{(x)}_\mu  \p^{(y)\mu}(\d(x-y) a_{\nu \c}(x)  a^\nu_{ \d}(x)) \[G^{\a\b\c\d} -G^{\b\a\c\d} \]=0, \eeq
\beq\label{d31}  \p^{(x)}_\mu  \p^{(y) }_\nu(\d(x-y) a^\mu_{ \c} (x) a^\nu_{ \d}(x)) \[G^{\a\c\b\d} -G^{\b\d\a\c}\]=0, \eeq
 \beq\label{d32} \p^\mu\d(x-y) (a_{\mu \d} a_{\nu \c}  a^\nu_{\eta}) (y)\[ f^{\b\d}_\u G^{\a\u\c\eta} +2f^{\b\c}_\u G^{\a\d\u\eta} -f^{\a\b}_\u G^{\u\d\c\eta}\]=0, \eeq
from which we see that $G^{\a\b\c\d} $ must be an invariant tensor on the Lie algebra ${\cal G}$, that  must be symmetric in its first (and last) two indices and it must be left invariant by the exchange of the first pair of indices with the second one. Therefore we have
$$K_G^\a(x)= {\d\o\d \u_\a(x)}{1\o4}\int dy\  G^{\b\c\d\epsilon} (a_{\mu \b}  a^\mu_{\c} a_{\nu \d}  a^\nu_{\epsilon})\equiv {\d\o\d \u_\a(x)} P.$$ Also $P$ can be added to $M$ in Equation (\ref{d24}).

  In this way we have proved Equation (\ref{81}) with $\int dx\ K(x)= M+N+P$ and $G_\a(x)$  satisfying Equations (\ref{85}),  (\ref{86}) and (\ref{87}).

   \end{document}